\documentclass[manuscript]{aastex631}

\shorttitle{ Seismic Minimum and Solar Magnetism Below the Surface}
\shortauthors{Jain et al.}

\begin{document}
\title{ What Seismic Minimum Reveals About Solar Magnetism Below the Surface?}
\correspondingauthor{Kiran Jain}
\email{kjain@nso.edu}

\author[0000-0002-1905-1639]{Kiran Jain}
\affiliation{National Solar Observatory,3665 Discovery Drive, Boulder, CO 80303, USA}

\author[0000-0002-1449-5417]{Niket Jain}
\affiliation{School of Computer Science and Engineering, Vellore Institute of Technology, Vellore, Tamilnadu 632014, India}

\author[0000-0002-4995-6180]{Sushanta C. Tripathy}
\affiliation{National Solar Observatory,3665 Discovery Drive, Boulder, CO 80303, USA}

\author[0000-0002-2227-0488]{Mausumi Dikpati}
\affiliation{High Altitude Observatory, NCAR, 3080 Center Green Drive, Boulder, CO 80301, USA}
%\altaffiltext{4}{Operated by the Association of Universities for Research in Astronomy under a cooperative agreement with the National Science Foundation. }

\begin{abstract}
The Sun's magnetic field varies in multiple time scales. Observations show that the minimum between cycles 24 and 25 was the second consecutive minimum which was deeper and wider than several earlier minima. Since the active regions observed at the Sun's surface are manifestations of the magnetic field generated in the interior, it is crucial to investigate/understand the dynamics below the surface. In this context, we report, by probing the solar interior with helioseismic techniques applied to long-term oscillations data from the Global Oscillation Network Group (GONG), that the seismic minima in deeper layers have been occurring about a year earlier than that at the surface for the last two consecutive solar cycles.  Our findings also demonstrate a decrease in strong magnetic fields at the base of the convection zone, the primary driver of the surface magnetic activity. We conclude that the magnetic fields located in the core and near-surface shear layers, in addition to the tachocline fields, play an important role in modifying the oscillation frequencies. This provides evidence that further strengthens the existence of a relic magnetic field in the Sun's core since its formation.
\end{abstract}


\section{INTRODUCTION}
\label{S-intro}
The Sun's magnetic field varies in multiple time scales  as evinced by the emergence and decay of sunspots on the solar surface. Among these timescales,  the 11-year cyclic pattern, commonly known as Solar Activity Cycle,  is the most prominent period.  Records of the solar activity measures such as sunspot number\footnote{\url{https://wwwbis.sidc.be/silso/datafiles}}, radio flux\footnote{\url{https://lasp.colorado.edu/lisird/data}} etc., reveal that the solar activity has been consistently, though gradually, declining for last several solar cycles. These observations also indicate that the last two activity minima are different from the earlier ones in recent times. These are much wider and deeper with more than 60\% spotless days around the cycle minimum. Furthermore, the galactic cosmic ray   modulation\footnote{\url{https://cosmicrays.oulu.fi}} that is anti-correlated with the solar activity also shows similar trends.  A close examination of the existing database of solar activity demonstrates that these type of minima were not unusual until Solar Cycle 19, the strongest cycle observed so far. In fact, the analysis of tree-ring data revealed an extended period of extremely low or no activity for several decades in the 17th and early 18th centuries, known as the Maunder Minimum  \citep{Eddy76}.  There were a few more periods of low activity in the past which are consistent with minima of the Centennial Gleissberg Cycle, a 90 -- 100 years variation observed on the Sun.  The recent period of  low activity is believed to be  among the minima  of Gleissberg Cycle that previously minimized at the beginning of the 20th century \citep{Feynman11}.

The activity minimum is a major milestone in the solar cycle when the magnetic field attains its lowest value before it starts increasing again. Low activity signifies the less probability of severe space weather events, e.g., flares, CMEs. This phase of the activity cycle is as important as the high activity phase since it provides a unique opportunity for studying the Sun's fundamental properties in the absence of strong magnetic fields. The magnetic field is generated in the solar interior and the dynamo responsible for 11-year cyclic activity in the Sun is believed to be seated near the base of the convection zone in a layer called {\it tachocline}. Thus, investigating the changes occurring in different layers beneath the solar surface  with the progression of the activity cycle is important. This can be achieved by analyzing the oscillation frequencies that vary in phase with the surface magnetic activity and exhibit a strong positive correlation with activity indicators \citep[e.g.,][]{Jain03,Jain09, Salabert15,Tripathy15, Simoniello16}.  Though all modes spend maximum time near the surface due to sharply increasing sound speed with depth, they still carry important information of the  layers they travel through and may provide crucial information about the magnetic field in the interior. With the availability of continuous oscillation data for more than 25 years,  it is now possible to study changes in more than two solar cycles. Similar to the Sun, there are  stellar activity cycles   in many stars with different periods  \citep{Garcia10,Chaplin14}, however the data is not as rich as for the Sun due to lack of long temporal coverage. Hence a better understanding of the physical processes taking place below the solar surface during  activity cycles is a path forward in asteroseismic studies. 

It is worth mentioning that the epochs of minimum preceding cycle 24 suggested different timings of minimum within the interior. For example, modes travelling down to the solar core indicated minimum about a year earlier than the reported activity minimum at the surface \citep{Salabert09} while the modes confined to the convection zone did not have any displacement in minimum  \citep{Tripathy10, jain11}. However, all these studies confirmed that the minimum between cycles 23 and 24 (MIN 23/24) was an extended and deeper minimum as compared to the minimum at the beginning of cycle 23 (MIN 22/23). These results were also supported by  numerous studies based on above the surface conditions. Despite observing another deep minimum,  not much attention has been paid so far on the recent minimum preceding cycle 25 (MIN 24/25). In this letter, we  examine the similarities and differences between the last two activity minima and investigate the possible scenario that might have led to two successive peculiar deep minima. 

\newpage
\section{ANALYSIS}
\label{S-data}
 In this work, we use $p$-mode frequencies for the individual ($n, \ell, m$) multiplets, $\nu_{n \ell m}$,  where $n$ is the radial order, $\ell$ is the harmonic degree and $m$ is the azimuthal order, running  from $-\ell$ to $+\ell$.  The mode frequencies for each multiplet were estimated from the $m-\nu$ power spectra constructed from the time series of individual 36-d period using   continuous Doppler observations from the Global Oscillation Network Group \citep[GONG;][]{harvey96,Jain21a}. Data analyzed here consist of 533 36-day overlapping data sets, with a spacing of 18 days between consecutive sets,  covering the period from 1995 May 7 to 2021 Aug 29 in 5-minute oscillation band in the frequency and degree ranges of  1860 $\le \nu \le$ 3450 $\mu$Hz and  0 $\le \ell \le$ 120, respectively. Thus, the analysis covers 3 consecutive  solar minima, starting about a year before the minimum of solar cycle 23 and  ending about a year after the minimum of solar cycle 25.  

We calculate the change in frequencies with reference to the guess frequencies that are used in the GONG {\it pmode} pipeline \citep{Hill96} for fitting  $\nu_{n \ell m}$.   Since the frequency shifts have well-known degree dependence \citep{Jain00}, the frequency shift for each mode is corrected for mode inertia \citep{JCD91}. In addition, the uncertainties in frequency determination are also included while computing the mean frequency shift, $\delta\nu$,  which is the error-weighted mean over all modes. We also compute weighted errors in $\delta\nu$ and these are shown in all figures. 
Finally, we take a running mean over 11 points to smooth out short-term variations (shown by solid red line in all figures) and use these smoothed shifts to define the epochs of minimum.  These epochs, known as the epochs of {\it seismic minimum}, are defined as the periods of lowest value of the smoothed frequency shifts within 1$\sigma$ errors. Since  $\delta\nu$ is known to vary with the change in solar activity  \citep{Jain09} and is best correlated with the 10.7 cm radio flux, $F_{10}$, we use radio flux as a proxy to determine the activity-related changes and the {\it activity minimum}.
 
 \section{RESULTS}
\label{S-results}

\begin{figure}
\epsscale{.85}
\plotone{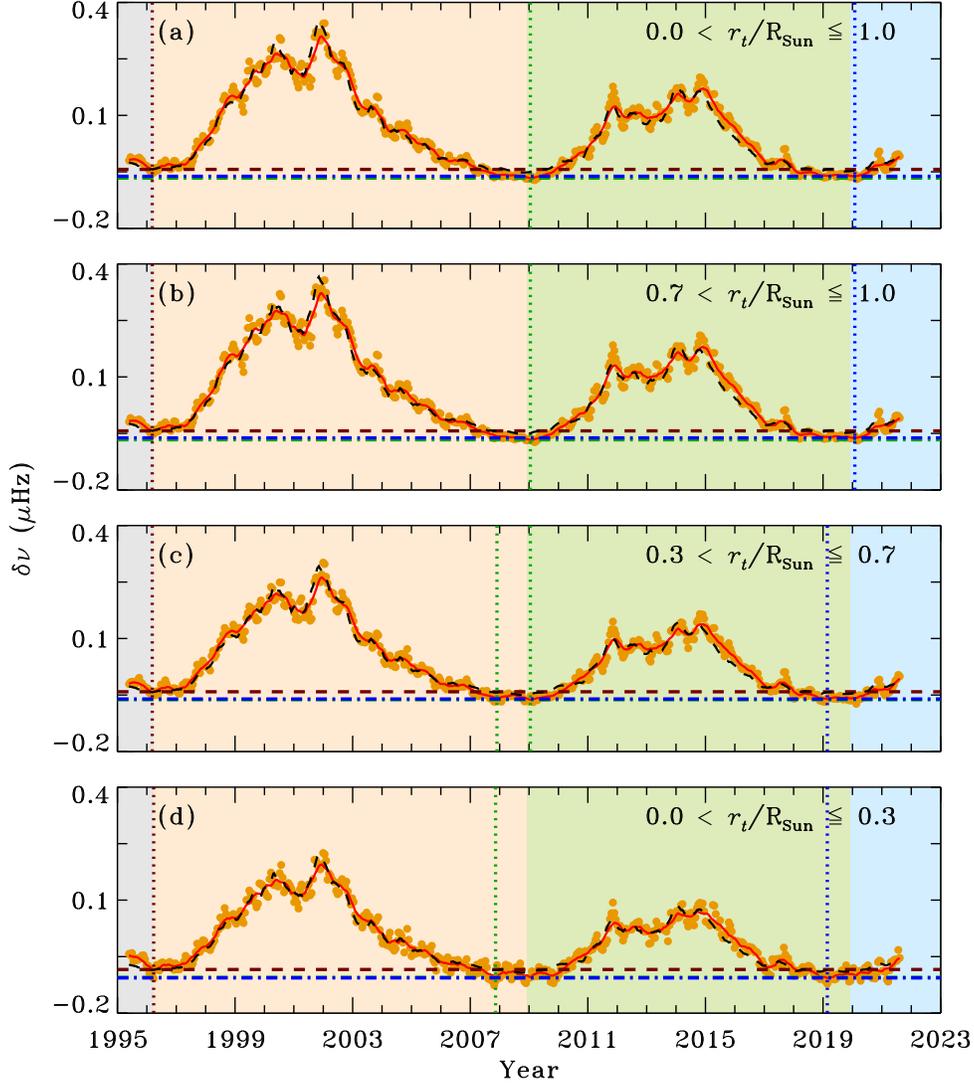}
\caption{Temporal evolution of  frequency shifts (symbols)
 for (a) all modes, and the modes with lower turning points in
(b) convection zone, (c) radiative zone, and (d) core. The errors in shifts are of the order of 10$^{-6}$ $\mu$Hz. An 11-point running mean of the frequency shifts is shown by the solid red line.  Vertical dotted lines show the epochs of minima in frequency shifts. The horizontal  lines represent the lowest values of the running mean during MIN 22/23 (brown), MIN 23/24 (green), and MIN 24/25 (blue).  The variations in smoothed radio flux are shown by black dashed line.   Background colors depict solar cycles (from left to right)  22 (partial), 23, 24, and 25 (partial) where the boundaries of different cycles are based on the activity minima.
\label{fig1}}
\end{figure}

Previous studies based on the helioseismic data suggest that the epochs of seismic minimum preceding cycle 24 varied with depth below the surface while no such depth dependence was found at MIN 22/23 \citep[e.g.,][]{jain11}. To investigate if these results were unique for MIN 23/24 only or a similar trend continued in the last minimum as well, the temporal variation of  $\delta\nu$ for different layers below the surface is displayed in Figure~\ref{fig1}. Top panel shows the variation of all modes covering the entire interior while lower  panels display variations in three distinct layers; 0.7 $< r_t/R_{\Sun} \le$ 1.0 (convection zone), 0.3 $< r_t/R_{\Sun} \le$ 0.7 (radiative zone), and 0.0 $< r_t/R_{\Sun} \le$ 0.3 (core) where $r_t$ is known as the lower-turning point radius defined by the following relation, 
 \begin{equation}
 r_t = \frac{c(r_t)}{2\pi}\frac{\sqrt{l(l+1)}}{\nu }, 
\end{equation}
here $c$ is the sound speed that increases sharply with depth. A higher value of  $\nu/\sqrt(l(l+1))$ denotes a smaller value of $r_t$, hence  a greater depth.
 
 As demonstrated in the figure,  the overall trends in all the panels are similar though the magnitudes are different. It is further seen that the amplitude of  $\delta\nu$ is higher in cycle 23 than in cycle 24. It is consistent with the solar-activity related variation  measured above the surface. In addition, the relative depths at all three minima reveal  that the oscillation frequencies during MIN 22/23 were not as low as these were during other two minima. Moreover, depths of the seismic minimum during MIN 23/24 and MIN 24/25 were comparable. The  over-plotted scaled variation of $F_{10}$, calculated over the same period as the frequency shifts and smoothed with an 11-point  running mean, also supports these inferences.  The difference in the depths of minimum during MIN 22/23 and MIN 23/24 was also reported earlier \citep[e.g.,][]{jain11,Broomhall17}  and our results are in agreement with those studies.   
 
It is further seen that all these epochs in the convection zone,  as identified in Figure~\ref{fig1}b, agreed with the activity minima. However, the seismic minima in lower two panels, displaying the changes in radiative zone and core, occurred about a year earlier than the activity minima during both MIN 23/24 and MIN 24/25. In fact, there was a double seismic minimum in the radiative zone that was earlier reported by \citet{jain11}. 
Further,  during the last two consecutive extended minima, the deeper layers experienced minimum earlier than the observed activity minimum at the surface. These findings clearly suggest a complex dynamical changes within the interior, and demonstrate that the changes occurring in deeper layers may not be consistent with those in the convection zone all of the time. This is not surprising as differential rotation patterns are the typical characteristics of convection zone while a rigid body rotation has been inferred in the radiative zone \citep{Thompson96}.
 
\begin{figure}
\vskip 1.5in
\epsscale{.85}
\plotone{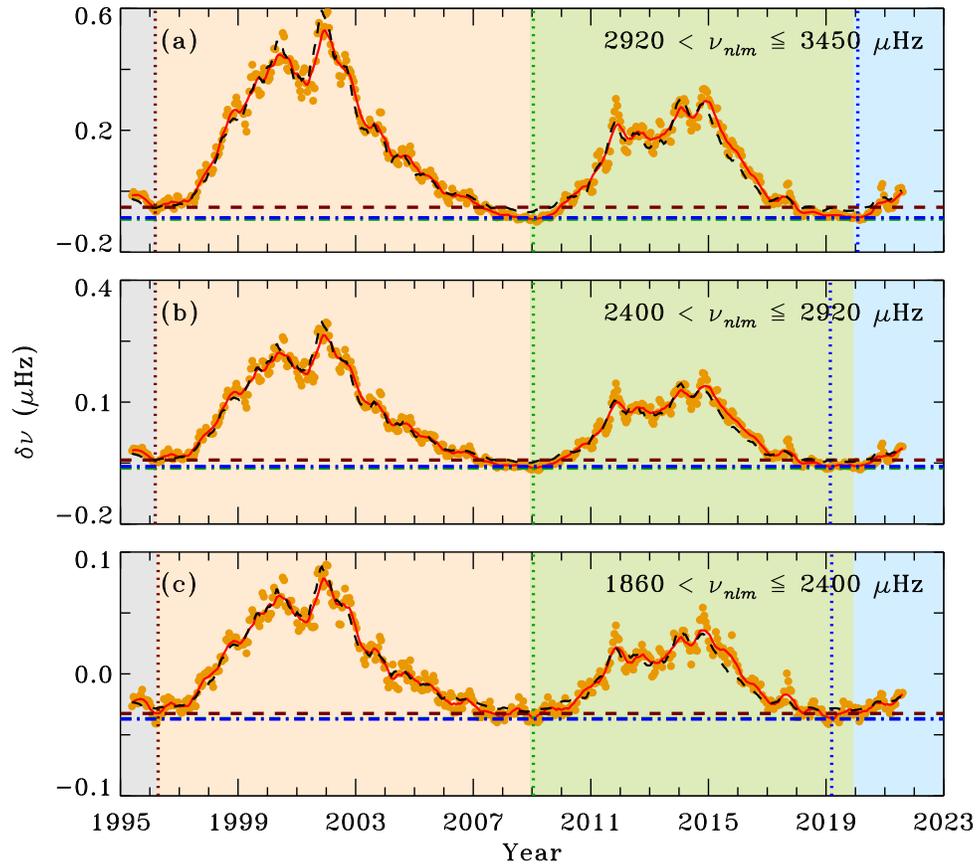}
\caption{Same as Figure~\ref{fig1}  but for modes with upper-turning points in (a) high-$\nu$, (b) mid-$\nu$ and (c) low-$\nu$ ranges. 
\label{fig2}}
\end{figure}

  We also examined the variability in near-surface layers that started around the maximum of cycle 23 as reported by \citet{Basu12}. For this, we divide entire data into three frequency bins:  2920 $\le \nu \le$ 3450 $\mu$Hz (high-$\nu$), 2400 $\le \nu \le$ 2920 $\mu$Hz (mid-$\nu$), and 1860 $\le \nu \le$ 2400 $\mu$Hz (low-$\nu$), and display  the temporal variation of  $\delta\nu$ in each frequency bin in Figure~\ref{fig2}.  Note that these frequency ranges are same as used by \citet{Basu12}  and  represent depth near the surface from which modes are reflected back into the Sun. This radius is generally referred to as the upper-turning point radius.  Modes in lower-$\nu$ range reflect back deeper inside the Sun than the other two ranges.  We find  that the seismic minima at MIN 22/23 in all frequency ranges occurred at the same time in early 1996  which also coincide with the activity minimum. Similarly, all  seismic minima  at MIN 23/24 also agreed with  the activity minimum.  However, during the latest minimum, i.e., MIN 24/25, we notice deviations in the epochs of seismic minimum from the activity minimum. While it  coincided with the activity minimum  (Figure~\ref{fig2}a) for high-$\nu$ range, the modes in mid- and low-$\nu$ ranges (Figures~\ref{fig2}b and ~\ref{fig2}c) sensed minima in early 2019 which is about a year prior to the activity minimum. This is puzzling since we find discrepancy in MIN 24/25 only, not in MIN 23/24 as reported by \citet{Basu12}. It is worth mentioning that the mode sets are very different in both studies. While modes used in this study provide information of the entire interior, \citet{Basu12} used  low-$\ell~(\le 3)$ modes from the integrated-light observations and the characteristics of these modes are sensitive to the core dynamics only, hence the comparison is not straightforward. 

\begin{figure}
\plotone{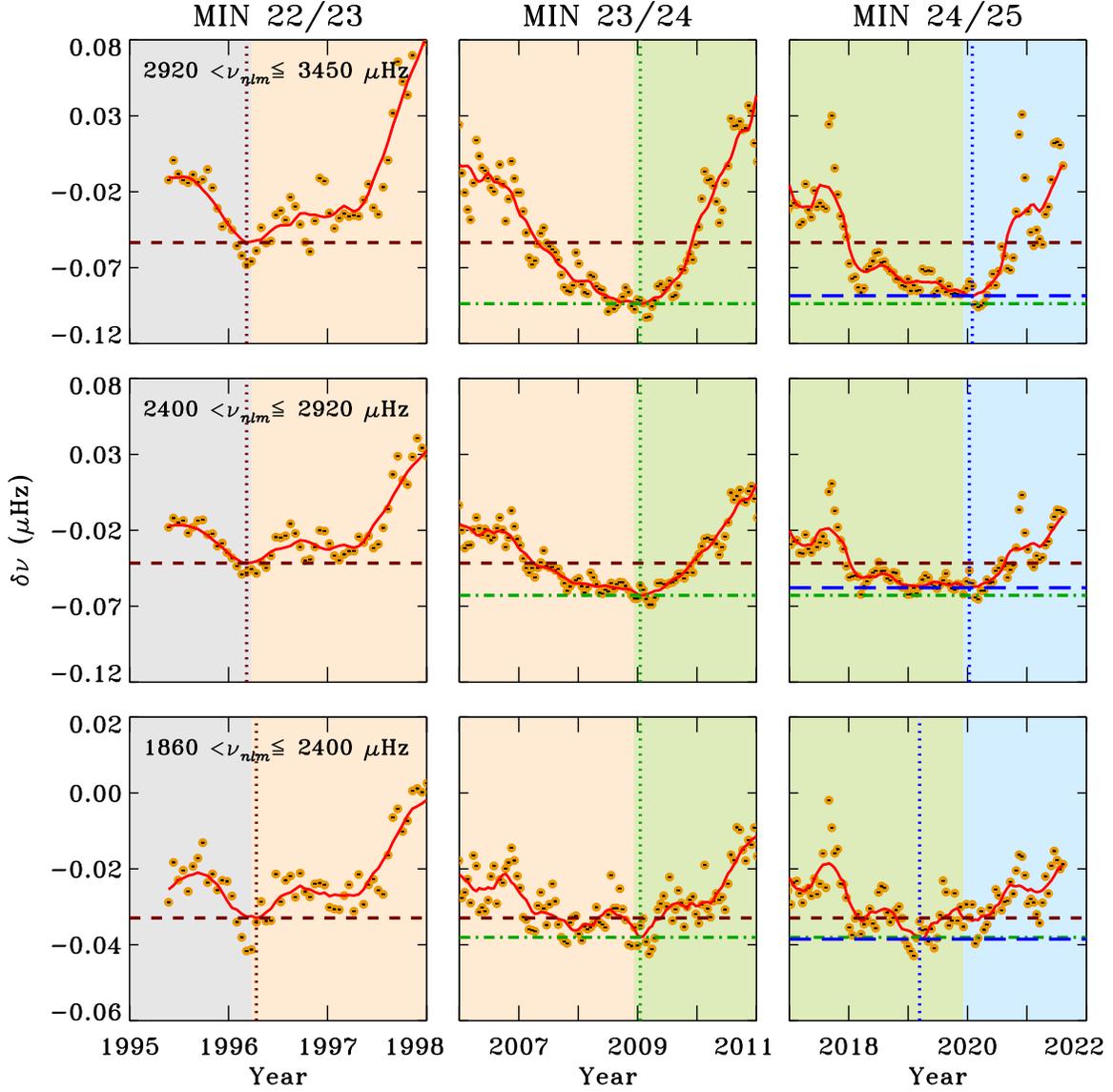}
\caption{Temporal evolution of  frequency shifts  (symbols) near 3 solar minima for the 
modes with lower turning points in  convection zone in three frequency ranges as marked in the leftmost panel of each row. The errors are smaller than the size of the symbols and shown by black. An 11-point running mean of the frequency shifts is shown by the solid red line. The horizontal  lines represent the lowest values of the running mean during MIN 22/23 (brown), MIN 23/24 (green), and MIN 24/25 (blue).  While boundaries of two background colors in each panel represent the activity minimum as marked on the top of each column, vertical dotted lines show the epochs of minima in frequency shifts.
\label{fig3}}
\end{figure}

\begin{figure}
\plotone{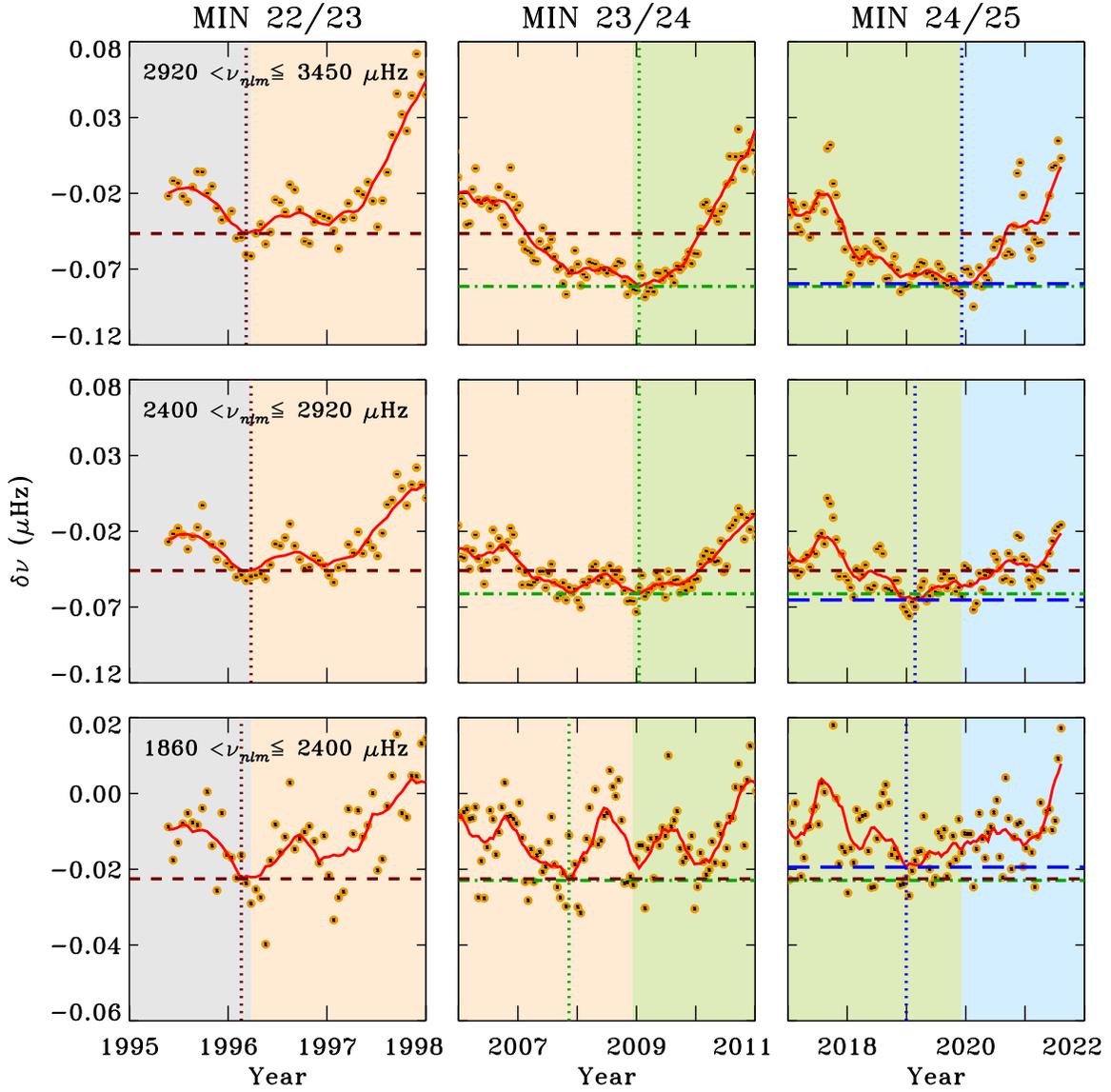}
\caption{Same as Figure~\ref{fig3} but for the modes with lower-turning points in the radiative zone.
\label{fig4}}
\end{figure}

\begin{figure}
\plotone{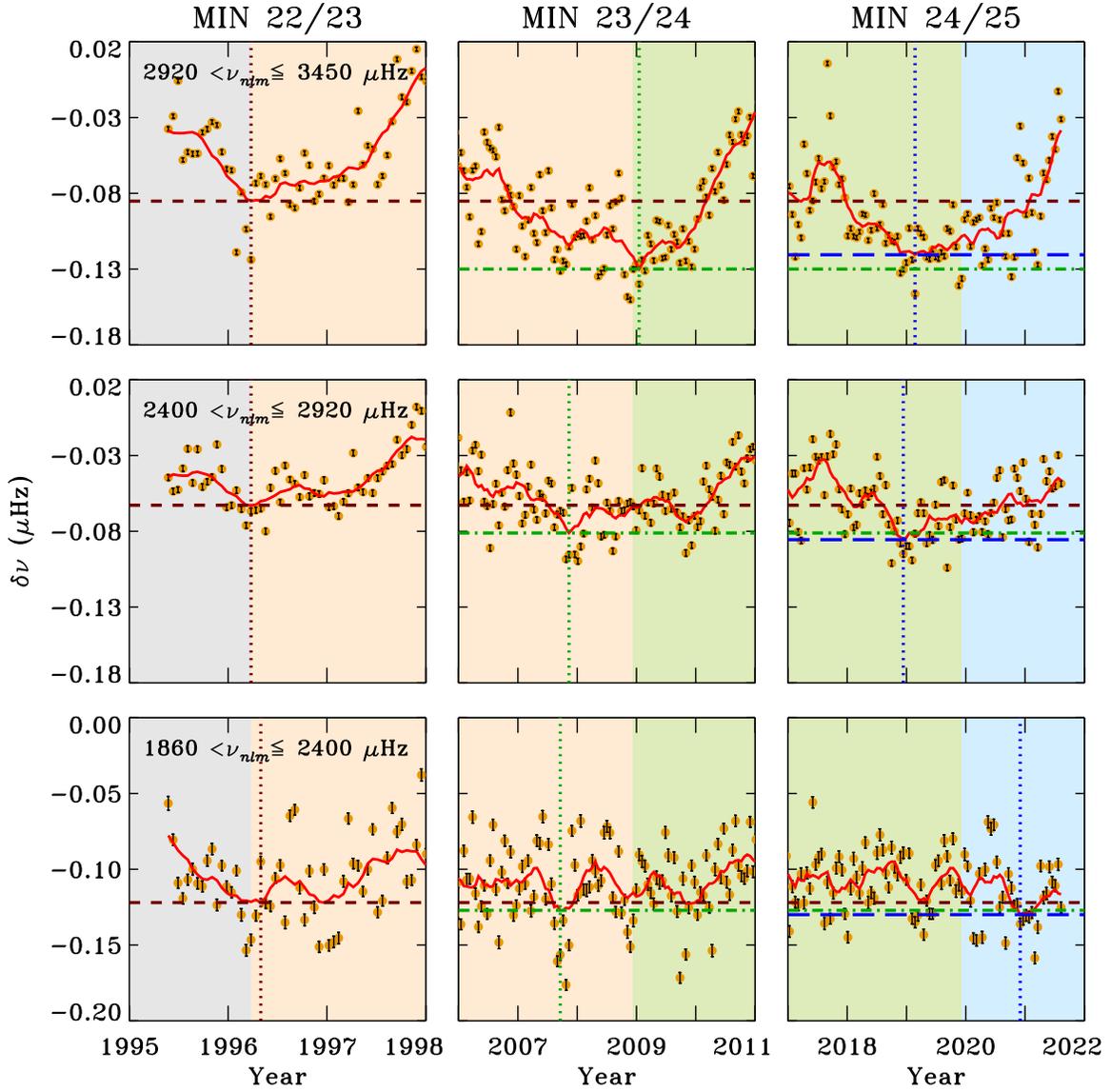}
\caption{Same as Figure~\ref{fig3} but for the modes with lower-turning points in core.
\label{fig5}}
\end{figure}

In order to understand the sources of this discrepancy, we further restrict modes on the basis of both lower- and upper-turning points.   For this purpose, we divide the entire data in 3 major groups, i.e., convection zone, radiative zone and the core, and each group is further divided into 3 sub-groups of frequency intervals described earlier. Results for each major group are presented in Figures~\ref{fig3} -- \ref{fig5}  where we display changes at low-activity periods of the solar minimum only. It is interesting to note again that there was no discrepancy between seismic and activity minima at MIN 22/23, however anomaly existed for other two minima which are discussed below.

For modes confined to the convection zone only, it is evident from Figure~\ref{fig3} that the seismic minima for modes with the upper-turning points in top two near-surface layers coincided with the activity minimum. However, modes that reflected back from the shallower layer (bottom row) exhibited seismic minimum in early 2019 contrary to the activity minimum in late 2019. For the modes with $r_t$ in  the radiative zone only, which means that all these modes carried additional information from the radiative zone, the  results are presented in Figure~\ref{fig4}. From  bottom two rows, we notice that the seismic minimum at MIN 24/25 occurred earlier than the activity minimum.  During MIN 23/24 shown in the middle column, both seismic and activity minima were in agreement  (middle row), there was a deviation for the low-frequency range (bottom row). This deviation is similar to that obtained for MIN 24/25  shown in Figure~\ref{fig3}. On further lowering $r_t$  to the core and using modes refracting back from this region only,  we find deviations in seismic minima during MIN 24/25 for all three frequency ranges  (Figure~\ref{fig5}). Note that the modes in lower two frequency ranges did sense an early seismic minimum during MIN 23/24 while these multiple epochs of minimum were not seen at MIN 22/23. It should be noted that the uncertainties in frequency determination tend to increase in deeper layers. Therefore,  we have used smoothed, error-weighted mean frequency shifts for robust determination of the epochs of seismic minimum.

Contrary to more than one epoch of seismic minimum  sensed by the different layers in deep interior as well as in near-surface layers during MIN 23/24 and MIN 24/25, the seismic minimum at MIN 22/23 was aligned with the activity minimum in all cases. This hints that the variability of magnetic fields below and above the convection zone are not ideally coupled.  In addition, the near-surface layers are also going through continual changes which support the argument put forward in several papers on the thinning down of  magnetic layers near the surface \citep{Basu12,Howe17, Jain18b}, these are not solely responsible for different epochs of the seismic minimum.   

\newpage
\section{DISCUSSION AND CONCLUSIONS}
\label{Summary}
It is evident that the layers from which a  propagating wave reflects back into the Sun or refracts towards the surface  have significant influence on the timing of seismic minimum.   Ever since the variability in oscillation frequencies has been discovered, a large number of studies confirmed that the magnetic field is primarily  responsible for changing frequencies. To explore the origin of the different timings of seismic minimum, we consider  three major sources of magnetic fields inside the Sun: {\it (i)}  a megagauss (strong) field located at tachocline that is primarily responsible for the dynamo mechanism \citep{Dziembowski89, Parker09} and the 11-year cyclic variation in solar activity, {\it (ii)} a weak field at the core that plays a crucial role in the generation of 22-year magnetic polarity cycle \citep{Sonett83, Mursula01} and is believed to have been present in the Sun since its formation \citep{Cowling45}, and {\it (iii)} another weak fields generated in NSSL, primarily due to small-scale dynamo action  but they are random and short-lived \citep{Vogler07}, can contribute too \citep{Broomhall12, Simoniello13a}. All three fields can play important roles in contributing to  the variation in oscillation frequencies and the multiple seismic minima -- the first two (namely items (i) and (ii)) in a systematic way, and the item (iii) in a more random way. Only a detailed modeling in the future can shed more lights on the relative contributions of these three fields.

Since the magnetic activity during MIN 22/23 was not weak compared to the other two minima, we postulate that this period was dominated by the strong fields generated in tachocline region while other two weak fields did not contribute significantly. As a result, both seismic and activity minima occurred around the same time irrespective of the regions where the modes travelled. However, the polar field strength decreased significantly in cycle 23 that led to a weak cycle 24. For the modes confined to outer 30\% of the interior with no influence from the region below tachocline,  NSSL remained magnetically strong during both MIN 22/23 and MIN 23/24, and exhibited  seismic and the activity minima  around the same time. However, this scenario changed during MIN 24/25 when the modes sampling the upper layers at depth approximately below 625 km (corresponding to 2400 $\mu$Hz) showed disagreement with the surface activity. This must have contributed to an early seismic minimum and suggest a disparity between magnetic fields located in tachocline and the near-surface layers.  Moreover,  fields generated in the tachocline were not strong enough to reduce the influence of weak fields in the near-surface shear layers. These inferences also support the findings of \citet{Kiefer18b} where authors suggested that the magnetic field affects the multiplet frequencies in a way that depends on the location and the geometry of the field inside the Sun. By comparing theoretical and  observed shifts, they concluded that strong tachocline fields cannot be sole responsible for the observed frequency shifts over the solar cycle and a  part of the surface effect in frequencies might be attributed to magnetic fields in the outer layers of the Sun. 

Furthermore, if the combination of tachocline and NSSL fields is alone responsible for changing the oscillation frequencies with time, similar results would have been obtained for all three depth zones. But the scenario is different for regions below the tachocline, particularly for modes with $r_t$ in the core where we find better agreement with \citet{Basu12}. This clearly hints that the fields located below tachocline also play important role in the dynamics of solar interior. The effect of relic (primordial) magnetic field was  already discussed in \citet{Gough90} and several subsequent studies but it was never fully understood due to lack of observational evidences. It is also believed that the relic field does not vary significantly with time and survives for thousands of years. As far as the cycle pattern is concerned, \citet{Dikpati06} showed that a regular cyclic dynamo can also be the origin of magnetic fields in the solar radiative tachocline and interior below.   There is a possibility that the cyclic pattern of these fields in conjunction with the primordial field senses a minimum earlier than the tachocline dynamo. During the extended periods of low activity, their influence is visible in the modes reflecting back from these regions. In addition,  these might be too weak compared to the magnetic field in tachocline that their influence is not recognized in stronger cycles. \citet{Sonett83} had also suggested that a steady poloidal field at or below the base of the convection zone would impart a bias to the solar cycle.

Since the continuous helioseismic observations sampling the entire interior are available only for last two solar cycles including 3 minima, such studies cannot be carried out for other weaker cycles and deep minima at the beginning 20th century.  Nevertheless, MIN 23/24 and MIN 24/25 are two consecutive minima that provide stronger evidence on the influence of relic magnetic field in the magnetism of deeper layers and finally on the solar oscillations. It would be interesting to track the conditions in cycle 25 and the minimum thereafter  for a better comprehension of dynamical changes in the solar interior. 

Data were acquired by GONG instruments operated by NISP/NSO/AURA/NSF.
%A part of this work was carried out by NJ for  his internship project. 
\facilities{GONG}

\bibliographystyle{aasjournal}

\begin{thebibliography}{}

\expandafter\ifx\csname natexlab\endcsname\relax\def\natexlab#1{#1}\fi
\providecommand{\url}[1]{\href{#1}{#1}}
\providecommand{\dodoi}[1]{doi:~\href{http://doi.org/#1}{\nolinkurl{#1}}}
\providecommand{\doeprint}[1]{\href{http://ascl.net/#1}{\nolinkurl{http://ascl.net/#1}}}
\providecommand{\doarXiv}[1]{\href{https://arxiv.org/abs/#1}{\nolinkurl{https://arxiv.org/abs/#1}}}

\bibitem[{{Basu} {et~al.}(2012){Basu}, {Broomhall}, {Chaplin}, \  {Elsworth}}]{Basu12} {Basu}, S., {Broomhall}, A.-M., {Chaplin}, W.~J., \& {Elsworth}, Y. 2012, \apj,  758, 43, \dodoi{10.1088/0004-637X/758/1/43}

\bibitem[{{Broomhall}(2017)}]{Broomhall17} {Broomhall}, A.~M. 2017, \solphys, 292, 67, \dodoi{10.1007/s11207-017-1068-5}

\bibitem[{{Broomhall} {et~al.}(2012){Broomhall}, {Chaplin}, {Elsworth}, \&  {Simoniello}}]{Broomhall12} {Broomhall}, A.~M., {Chaplin}, W.~J., {Elsworth}, Y., \& {Simoniello}, R. 2012,  \mnras, 420, 1405, \dodoi{10.1111/j.1365-2966.2011.20123.x}

\bibitem[{{Chaplin} \& {Basu}(2014)}]{Chaplin14} {Chaplin}, W.~J., \& {Basu}, S. 2014, \ssr, 186, 437,  \dodoi{10.1007/s11214-014-0090-2}

\bibitem[{{Christensen-Dalsgaard} \& {Berthomieu}(1991)}]{JCD91} {Christensen-Dalsgaard}, J., \& {Berthomieu}, G. 1991, {Theory of solar  oscillations.} (Tucson: University of Arizona Press), 401--478

\bibitem[{{Cowling}(1945)}]{Cowling45} {Cowling}, T.~G. 1945, \mnras, 105, 166, \dodoi{10.1093/mnras/105.3.166}

\bibitem[{{Dikpati} {et~al.}(2006){Dikpati}, {Gilman}, \&  {MacGregor}}]{Dikpati06} {Dikpati}, M., {Gilman}, P.~A., \& {MacGregor}, K.~B. 2006, \apj, 638, 564,  \dodoi{10.1086/498682}

\bibitem[{{Dziembowski} \& {Goode}(1989)}]{Dziembowski89} {Dziembowski}, W.~A., \& {Goode}, P.~R. 1989, \apj, 347, 540,  \dodoi{10.1086/168144}

\bibitem[{{Eddy}(1976)}]{Eddy76} {Eddy}, J.~A. 1976, Science, 192, 1189, \dodoi{10.1126/science.192.4245.1189}

\bibitem[{{Feynman} \& {Ruzmaikin}(2011)}]{Feynman11} {Feynman}, J., \& {Ruzmaikin}, A. 2011, \solphys, 272, 351,  \dodoi{10.1007/s11207-011-9828-0}

\bibitem[{{Garc{\'\i}a} {et~al.}(2010){Garc{\'\i}a}, {Mathur}, {Salabert},  {Ballot}, {R{\'e}gulo}, {Metcalfe}, \& {Baglin}}]{Garcia10} {Garc{\'\i}a}, R.~A., {Mathur}, S., {Salabert}, D., {et~al.} 2010, Science, 329, 1032, \dodoi{10.1126/science.1191064}

\bibitem[{{Gough} \& {Thompson}(1990)}]{Gough90} {Gough}, D.~O., \& {Thompson}, M.~J. 1990, \mnras, 242, 25,  \dodoi{10.1093/mnras/242.1.25}

\bibitem[{{Harvey} {et~al.}(1996){Harvey}, {Hill}, {Hubbard}, {Kennedy},   {Leibacher}, {Pintar}, {Gilman}, {Noyes}, {Title}, {Toomre}, {Ulrich},  {Bhatnagar}, {Kennewell}, {Marquette}, {Patron}, {Saa}, \& {Yasukawa}}]{harvey96} {Harvey}, J.~W., {Hill}, F., {Hubbard}, R.~P., {et~al.} 1996, Science, 272, 1284, \dodoi{10.1126/science.272.5266.1284}

\bibitem[{{Hill} {et~al.}(1996){Hill}, {Stark}, {Stebbins}, {Anderson},  {Antia}, {Brown}, {Duvall}, {Haber}, {Harvey}, {Hathaway}, {Howe}, {Hubbard},  {Jones}, {Kennedy}, {Korzennik}, {Kosovichev}, {Leibacher}, {Libbrecht},  {Pintar}, {Rhodes}, {Schou}, {Thompson}, {Tomczyk}, {Toner}, {Toussaint}, \&  {Williams}}]{Hill96} {Hill}, F., {Stark}, P.~B., {Stebbins}, R.~T., {et~al.} 1996, Science, 272,  1292, \dodoi{10.1126/science.272.5266.1292}

\bibitem[{{Howe} {et~al.}(2017){Howe}, {Davies}, {Chaplin}, {Elsworth}, {Basu},  {Hale}, {Ball}, \& {Komm}}]{Howe17} {Howe}, R., {Davies}, G.~R., {Chaplin}, W.~J., {et~al.} 2017, \mnras, 470,   1935, \dodoi{10.1093/mnras/stx1318}

\bibitem[{{Jain} \& {Bhatnagar}(2003)}]{Jain03} {Jain}, K., \& {Bhatnagar}, A. 2003, \solphys, 213, 257,  \dodoi{10.1023/A:1023923817099}

\bibitem[{{Jain} {et~al.}(2018){Jain}, {Tripathy}, {Hill}, {Salabert},  {Garc{\'{\i}}a}, \& {Broomhall}}]{Jain18b} {Jain}, K., {Tripathy}, S., {Hill}, F., {et~al.} 2018, in IAU Symposium, Vol.  340, IAU Symposium, ed. D.~{Banerjee}, J.~{Jiang}, K.~{Kusano}, \&  S.~{Solanki}, 27--30, \dodoi{10.1017/S174392131800162X}

\bibitem[{{Jain} {et~al.}(2000){Jain}, {Tripathy}, \& {Bhatnagar}}]{Jain00} {Jain}, K., {Tripathy}, S.~C., \& {Bhatnagar}, A. 2000, \apj, 542, 521,   \dodoi{10.1086/309508}

\bibitem[{{Jain} {et~al.}(2009){Jain}, {Tripathy}, \& {Hill}}]{Jain09} {Jain}, K., {Tripathy}, S.~C., \& {Hill}, F. 2009, \apj, 695, 1567, \dodoi{10.1088/0004-637X/695/2/1567}

\bibitem[{{Jain} {et~al.}(2011){Jain}, {Tripathy}, \& {Hill}}]{jain11} ---. 2011, \apj, 739, 6, \dodoi{10.1088/0004-637X/739/1/6}

\bibitem[{{Jain} {et~al.}(2021){Jain}, {Tripathy}, {Hill}, \&  {Pevtsov}}]{Jain21a} {Jain}, K., {Tripathy}, S.~C., {Hill}, F., \& {Pevtsov}, A.~A. 2021, \pasp,   133, 105001, \dodoi{10.1088/1538-3873/ac24d5}

\bibitem[{{Kiefer} \& {Roth}(2018)}]{Kiefer18b} {Kiefer}, R., \& {Roth}, M. 2018, \apj, 854, 74,   \dodoi{10.3847/1538-4357/aaa3f7}

\bibitem[{{Mursula} {et~al.}(2001){Mursula}, {Usoskin}, \&  {Kovaltsov}}]{Mursula01} {Mursula}, K., {Usoskin}, I.~G., \& {Kovaltsov}, G.~A. 2001, \solphys, 198, 51, \dodoi{10.1023/A:1005218414790}

\bibitem[{{Parker}(2009)}]{Parker09} {Parker}, E.~N. 2009, \ssr, 144, 15, \dodoi{10.1007/s11214-008-9445-x}

\bibitem[{{Salabert} {et~al.}(2009){Salabert}, {Garc{\'{\i}}a}, {Pall{\'e}}, \&  {Jim{\'e}nez-Reyes}}]{Salabert09} {Salabert}, D., {Garc{\'{\i}}a}, R.~A., {Pall{\'e}}, P.~L., \&   {Jim{\'e}nez-Reyes}, S.~J. 2009, \aap, 504, L1, \dodoi{10.1051/0004-6361/200912736}

\bibitem[{{Salabert} {et~al.}(2015){Salabert}, {Garc{\'{\i}}a}, \& {Turck-Chi{\`e}ze}}]{Salabert15} {Salabert}, D., {Garc{\'{\i}}a}, R.~A., \& {Turck-Chi{\`e}ze}, S. 2015, \aap,  578, A137, \dodoi{10.1051/0004-6361/201425236}

\bibitem[{{Simoniello} {et~al.}(2013){Simoniello}, {Jain}, {Tripathy}, {Turck-Chi{\`e}ze}, {Baldner}, {Finsterle}, {Hill}, \&  {Roth}}]{Simoniello13a} {Simoniello}, R., {Jain}, K., {Tripathy}, S.~C., {et~al.} 2013, \apj, 765, 100, \dodoi{10.1088/0004-637X/765/2/100}

\bibitem[{{Simoniello} {et~al.}(2016){Simoniello}, {Tripathy}, {Jain}, \&  {Hill}}]{Simoniello16} {Simoniello}, R., {Tripathy}, S.~C., {Jain}, K., \& {Hill}, F. 2016, \apj, 828,  41, \dodoi{10.3847/0004-637X/828/1/41}

\bibitem[{{Sonett}(1983)}]{Sonett83} {Sonett}, C.~P. 1983, \nat, 306, 670, \dodoi{10.1038/306670a0}

\bibitem[{{Thompson} {et~al.}(1996){Thompson}, {Toomre}, {Anderson}, {Antia},  {Berthomieu}, {Burtonclay}, {Chitre}, {Christensen-Dalsgaard}, {Corbard}, {De  Rosa}, {Genovese}, {Gough}, {Haber}, {Harvey}, {Hill}, {Howe}, {Korzennik}, {Kosovichev}, {Leibacher}, {Pijpers}, {Provost}, {Rhodes}, {Schou}, {Sekii},  {Stark}, \& {Wilson}}]{Thompson96} {Thompson}, M.~J., {Toomre}, J., {Anderson}, E.~R., {et~al.} 1996, Science, 272, 1300, \dodoi{10.1126/science.272.5266.1300}

\bibitem[{{Tripathy} {et~al.}(2015){Tripathy}, {Jain}, \& {Hill}}]{Tripathy15} {Tripathy}, S.~C., {Jain}, K., \& {Hill}, F. 2015, \apj, 812, 20, \dodoi{10.1088/0004-637X/812/1/20}

\bibitem[{{Tripathy} {et~al.}(2010){Tripathy}, {Jain}, {Hill}, \&  {Leibacher}}]{Tripathy10} {Tripathy}, S.~C., {Jain}, K., {Hill}, F., \& {Leibacher}, J.~W. 2010, \apjl,  711, L84, \dodoi{10.1088/2041-8205/711/2/L84}

\bibitem[{{V{\"o}gler} \& {Sch{\"u}ssler}(2007)}]{Vogler07} {V{\"o}gler}, A., \& {Sch{\"u}ssler}, M. 2007, \aap, 465, L43, \dodoi{10.1051/0004-6361:20077253}

\end{thebibliography}

\clearpage

\clearpage

\end{document}